\documentclass[aip,graphicx]{revtex4-1}
\usepackage{graphicx}

\draft 

\begin{document}


\title{Anomalous metamagnetic-like transition in a FeRh/Fe$_3$Pt interface occurring at T $\approx$ 120 K in the field-cooled-cooling curves for low magnetic fields.} 
\author{S. Salem-Sugui Jr.}
\affiliation{Instituto de Fisica, Universidade Federal do Rio de Janeiro,
21941-972 Rio de Janeiro, RJ, Brazil}
\author{A. D. Alvarenga}
\affiliation{Instituto Nacional de Metrologia, Qualidade e Tecnologia, 25250-020 Duque de Caxias, RJ, Brazil.}
\author{R.D. Noce}
\affiliation{Instituto de Quimica, UNESP, 1801-400, Araraquara, SP, Brazil.} 
\author{R.B. Guimar\~aes}
\affiliation{Instituto de Fisica, Universidade Federal Fluminense,  24210-346, Niteroi, RJ, Brazil}
\author{C. Salazar Mejia}
\affiliation{Instituto de Fisica, Universidade Federal do Rio de Janeiro,
21941-972 Rio de Janeiro, RJ, Brazil}
\author{H. Salim}
\affiliation{Instituto de Fisica, Universidade Federal do Rio de Janeiro,
21941-972 Rio de Janeiro, RJ, Brazil}
\author{F.G. Gandra}
\affiliation{Instituto de Fisica, UNICAMP, Campinas, SP, Brazil}
\date{\today}
\begin{abstract}
We report on the magnetic properties of a special configuration of a FeRh thin film. An anomalous behavior on the magnetisation vs. temperature was observed when low magnetic fields are applied in the plane of a thin layer of FeRh deposited on ordered Fe$_3$Pt. The anomalous effect resembles a metamagnetic transition and occur only in the field-cooled-cooling magnetisation curve at temperatures near 120 K in samples without any heat treatment.  
\end{abstract}

\maketitle 
The Fe$_{1-x}$Rh$_x$ system with $x$ $\approx$ 0.5 crystalize in the CsCl structure under certain heat-treatment conditions \cite{k1,k2,k3}. In this cubic phase, equiatomic FeRh is antiferromagnetic and exhibits a metamagnetic magnetostructural phase transition at ambient temperature. The transition temperature is however highly sensitive to small changes out of the  equiatomic stoichiometry \cite{kuba} and microstructural scale \cite{kang,marquina}. The magnetostructural phase transition is of first order kind with a volume expansion of about 1$\%$ when entering in the ferromagnetic phase and a temperature hysteresis of the order of 10 K.\cite{k1,maat,ibarra,anna} These characteristics make thin films of the FeRh system to present a good potential for applications as micro-electromechanical devices \cite{thiele}. The study which motivated us, performed in the binary thin film of FePt-FeRh presented in Ref.\onlinecite{thiele}, is, in particular, very interesting.

In this work, we study Fe$_{1-x}$Rh$_x$ in the composition $x$ $\approx 0.5$ as obtained by electrodeposition on a foil of ordered Fe$_3$Pt.  This work addresses the possibility of obtaining a thin film of FeRh in the antiferromagnetic phase by electrodeposition on a foil of a ferromagnetic compound, forming a self-sustained film. We chose the ferromagnetic systems to be ordered Fe$_3$Pt, which is a well known and studied ferromagnet \cite{prl,prl2}, for which the preparation of a thin foil is quite simple\cite{ercan}. The preparation and study of such a self-sustained film stack, searching for FeRh in the antiferromagnetic phase, was our main motivation. We report here the first results obtained in the film stack Fe$_3$Pt-FeRh prior to any heat treatment. The system as prepared, presents an anomaly near 120 K in the temperature dependent magnetisation curve when cooled from 300 K, in very low applied magnetic fields. The anomaly appears for fields as low as 20 Oe and it is reproducible, resembling a metamagnetic transition. 
 
The Fe$_{1-x}$Rh$_x$ with $x$ $\approx 0.5$ thin film was obtained by electrodeposition on a 0.6 cm$^2$ of a thick ordered Fe$_3$Pt film.  The electrodeposition occurred in galvanostatic conditions from a bath composed of: 0.01 Molar Fe$_2$(SO$_4$)$_3$, 0.001- 0.0001 Molar Rh$_2$(SO$_4$)$_3$, 0.1 Molar K$_2$SO$_4$ and 0.05 Molar Sodium citrate (Na$_3$C$_6$H$_5$O$_7$). The final film thickness of FeRh falls in the range of 30-50 nm, depending on the applied current density,  which was varied from 1 to 5 mAcm$^{-2}$, and the charge for preparing all deposits was kept to 10 C.\cite{noce,taba,schulz} An EG-G PAR potentiostat-galvanostat, model 273, served as a constant current source.  The electrodeposition occur under previously established conditions  producing an approximately equiatomic deposition of Fe and Rh.\cite{noce} The Fe$_3$Pt substrate with $\approx$ 3 $\mu$m thickness was obtained by cold-rolling a previously prepared arc-melted sample. The ordered phase of the Fe$_3$Pt foil was achieved after an appropriated heat treatment as described in Ref. \onlinecite{ercan}.  A foil with 0.6x0.5 cm$^2$ ($m$=0.00271 g) of the final film was used in the measurements.  

Figure 1 shows a low angle (2 degrees) X-ray diffraction (XRD) curve obtained from the FeRh surface of the film and the inset shows usual XRD curves obtained in the same FeRh surface film and in the Fe$_3$Pt foil used as substrate. It is important to mention that the FeRh structure could not be resolved by means of usual XRD, as shown in the inset of Fig. 1. The double plot in the inset of Fig.1 evidences  the one-to-one match of the peaks of the ordered Fe$_3$Pt cubic structure in both diffraction curves. The low angle XRD data was refined for the peaks indicated in Fig. 1, following the Rietveld method with the MAUD software\cite{Lutterotti} and  using ICSD data base. The low angle diffraction curve was obtained about one year after \cite{tech} the diffraction curve presented in the inset of Fig. 1 was obtained and, besides the Fe$_3$Pt peaks, it also shows a signal due to Fe$_3$O$_4$, which is absent in the diffraction curve show in the inset figure. This fact evidences that Fe$_3$O$_4$ grew in the FeRh surface of the film. It should be mentioned that Energy Dispersive X-ray Spectrometry performed on many Fe$_{1-x}$Rh$_x$ thin films grown by electrodeposition under same conditions as here did not show oxygen peaks.\cite{noce}  Based on that, one may speculate that Fe$_3$O$_4$ grow on the FeRh surface as a result of surface oxidation. Despite that, it was possible to identify small peaks in the low angle diffraction curve belonging to the FeRh cubic structure (CsCl type) with lattice parameter a = 2.98 \AA\ , as calculated from the FeRh peaks indicated in Fig. 1. 

Magnetisation data were obtained by using a PPMS-9T Quantum-Design magnetometer. All data were obtained with the magnetic field applied parallel to the plane of the films. Isofield curves, $M vs. T$, and isothermal curves, $M vs. H$,  were all obtained after cooling the studied sample from 300 K, in zero applied  magnetic field (zfc), but in the presence of the earth«s magnetic field, to a desired temperature. After that, for the isofield $M vs. T$ curves, a magnetic field was applied reaching the desired value without overshooting. The zfc (zero-field-cooled) data was then collected by heating the sample at fixed increments of temperature up to 300 K, which was followed by the fcc (field-cooled-cooling) sequence with same $\delta T$ increments down to 1.8 K. Finally, a fch (field-cooled-heating) process ends the sequence at 300 K. These procedures correspond to a cycle starting with a zfc (zero-field-cooled) curve obtained by heating from 1.8 K up to 300 K followed by a fcc (field-cooled-cooling) curve  followed by a fch (field-cooled-heating) respectively. These data were obtained for fields ranging from 20 Oe to 10 kOe. For isothermals $M vs. H$ curves, magnetisation is measured as the magnetic field increases (or decreases depending on the branch of the hysteresis curve) at fixed increments.

\begin{figure}
\includegraphics[width=\linewidth]{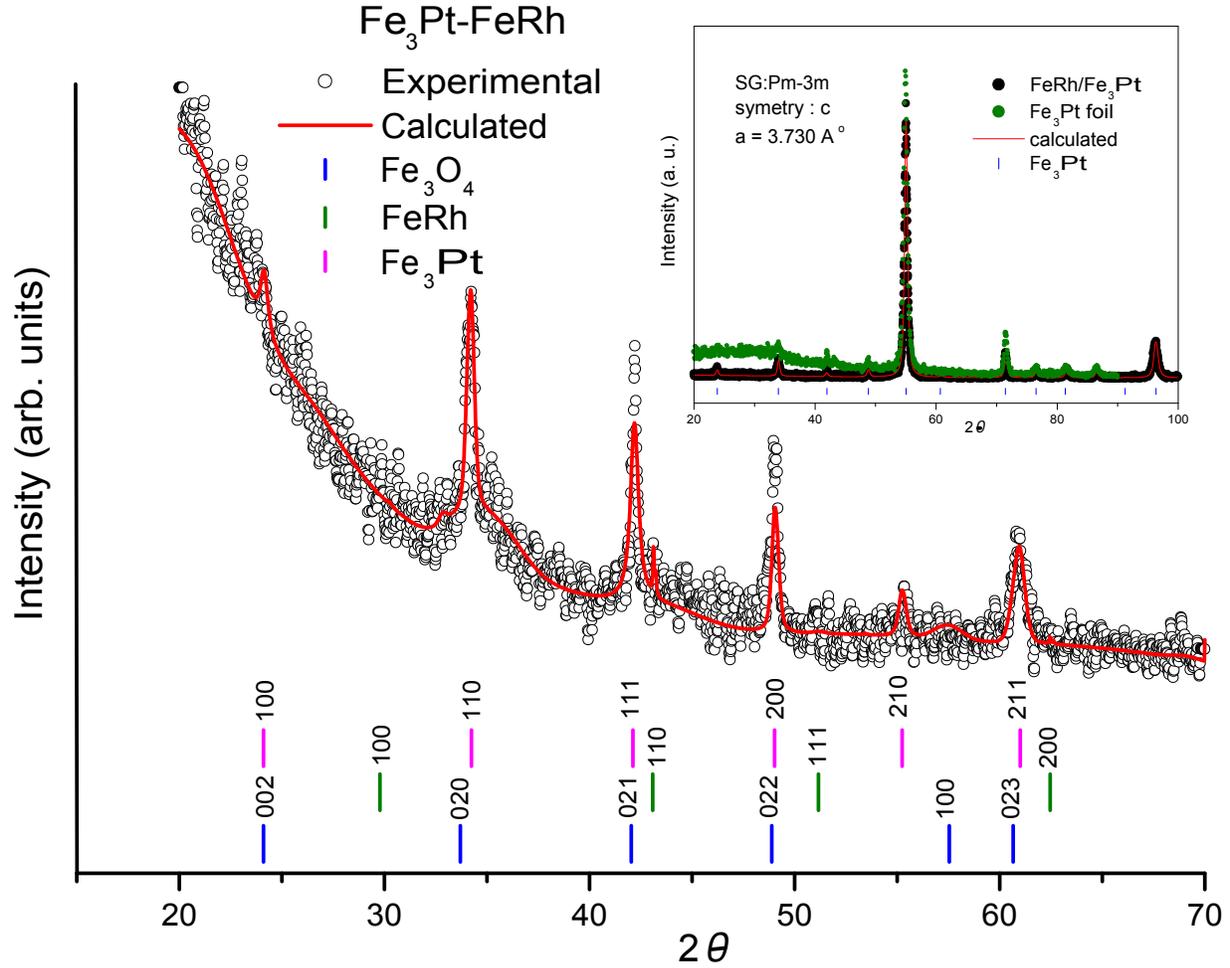}
\caption{Low angle XRD of the FeRh face.  The Inset  shows X-ray diffractions of Fe$_3$Pt and FeRh-Fe$_3$Pt  films. Perpendicular bars appearing below XRD in the main figure and its inset represent the peaks resolved for each correspondent structure.}
\label{fig1}
\end{figure} 
\begin{figure}
\includegraphics[width=\linewidth]{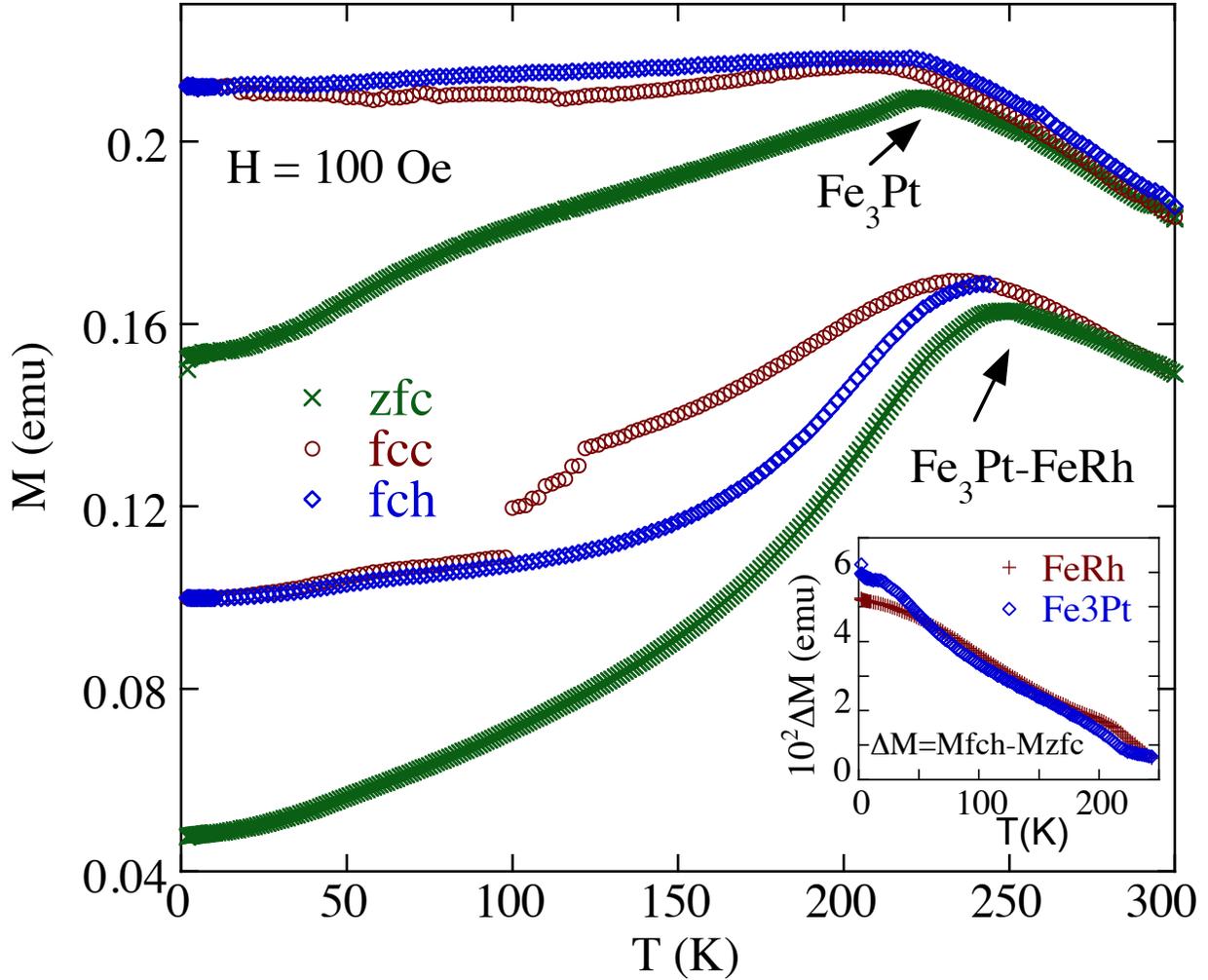}
\caption{ Magnetisation curves for H = 100 Oe for Fe$_3$Pt and FeRh-Fe$_3$Pt films. The inset shows a plot of M$_{fch}$-M$_{zfc}$ vs. T as obtained for each compound. }
\label{fig2}
\end{figure}

Figure 2 displays the $M vs. T$ curves obtained with a magnetic field $H$ = 100 Oe applied on a pure Fe$_3$Pt thin foil and on the film stack Fe$_{1-x}$Rh$_x$-Fe$_3$Pt  (both films with approximately same mass). The curves of each system are shifted along the Y-axis for better presentation. Figure 2 allows the visualization of the overall effect of the thin FeRh layer deposited on the Fe$_3$Pt foil. It is possible to see that the fcc and fch curves in Fe$_3$Pt are much closer below 200 K than in the film stack. For FeRh-Fe$_3$Pt, a separation between the fcc and fch curves starts below $\approx$ 220 K, followed by an anomaly in the fcc curve appearing as temperature drops below $\approx$ 120 K, with a jump at 100 K. It is interesting to observe that both fcc and fch curves for FeRh-Fe$_3$Pt virtually join together below the jump in the magnetisation at 100 K.  One may interpret the anomaly in the fcc curve of the film stack as a metamagnetic transition, similar to that observed in pure FeRh but with  a composition out of the equiatomic stoichiometry \cite{kuba}.  However, the later can not explain the fact that the fch curve of the  film stack evolves continuously as T increases above 100 K, with values lying well below the fcc curve (this up to temperatures near 220 K). As mentioned, the fcc and fch curves for pure Fe$_3$Pt are much closer when compared to the large separation of the same curves in the film stack, evidencing the effect of the FeRh layer in the fch curve in the temperature region 100 K - 220 K. The same effect seems to explain the reduction in the 
magnetisation of the zfc curve of the film stack relatively to the zfc curve of Fe$_3$Pt (for Fe$_3$Pt, the reduction in magnetization observed in the zfc curve below 230 K is a consequence of a misalignment of the domains, since as the temperature (or field) is increased the domains become more aligned). The zfc curve of the film stack, which lies well below the correspondent fch curve, also evolves continuously from  1.8 K up to 220 
K. To better exemplify this discussion, it is plotted in the inset of Fig. 2 results of the subtraction between the fch and zfc magnetisation curves of Fe$_3$Pt, and between the fch and zfc curves of FeRh-Fe$_3$Pt (both from Fig. 2). As it is possible to see in this inset, the resulting $\Delta M$ = $M_{fch}$-$M_{zfc}$ of each system show quite similar values and behavior with temperature, highly suggesting that the effect which reduces the fch magnetisation in the FeRh-Fe$_3$Pt film, also reduces the correspondent zfc magnetisation. These facts may suggest that the deposited FeRh layer has an approximately equiatomic stoichiometry which, if isolated, would show the metamagnetic transition near 220 K. In that case, the anomaly in the fcc curve observed near 120 K when cooling the film in a low magnetic field, as well the continuously increasing of magnetisation observed in the fch curve (and in the zfc curve) when heating the sample from temperatures below the "jump", might be explained, possibly, in terms of the magnetic coupling between the ferromagnetic Fe$_3$Pt bulk and the antiferromagnetic thin layer of FeRh. Within this scenarios, the differences between the Fe$_3$Pt and FeRh-Fe$_3$Pt curves of Fig. 2 are due to the exchange antiferromagnetic-ferromagnetic coupling, which favor a metamagnetic-like transition occurring near 120 K, only when cooling the film in a fcc curve. It is also interesting to note in Fig. 2 that, contrary to what occurs for FeRh-Fe$_3$Pt, the fch curve for Fe$_3$Pt lies somewhat above the fcc curve.

\begin{figure}[t]
\includegraphics[width=\linewidth]{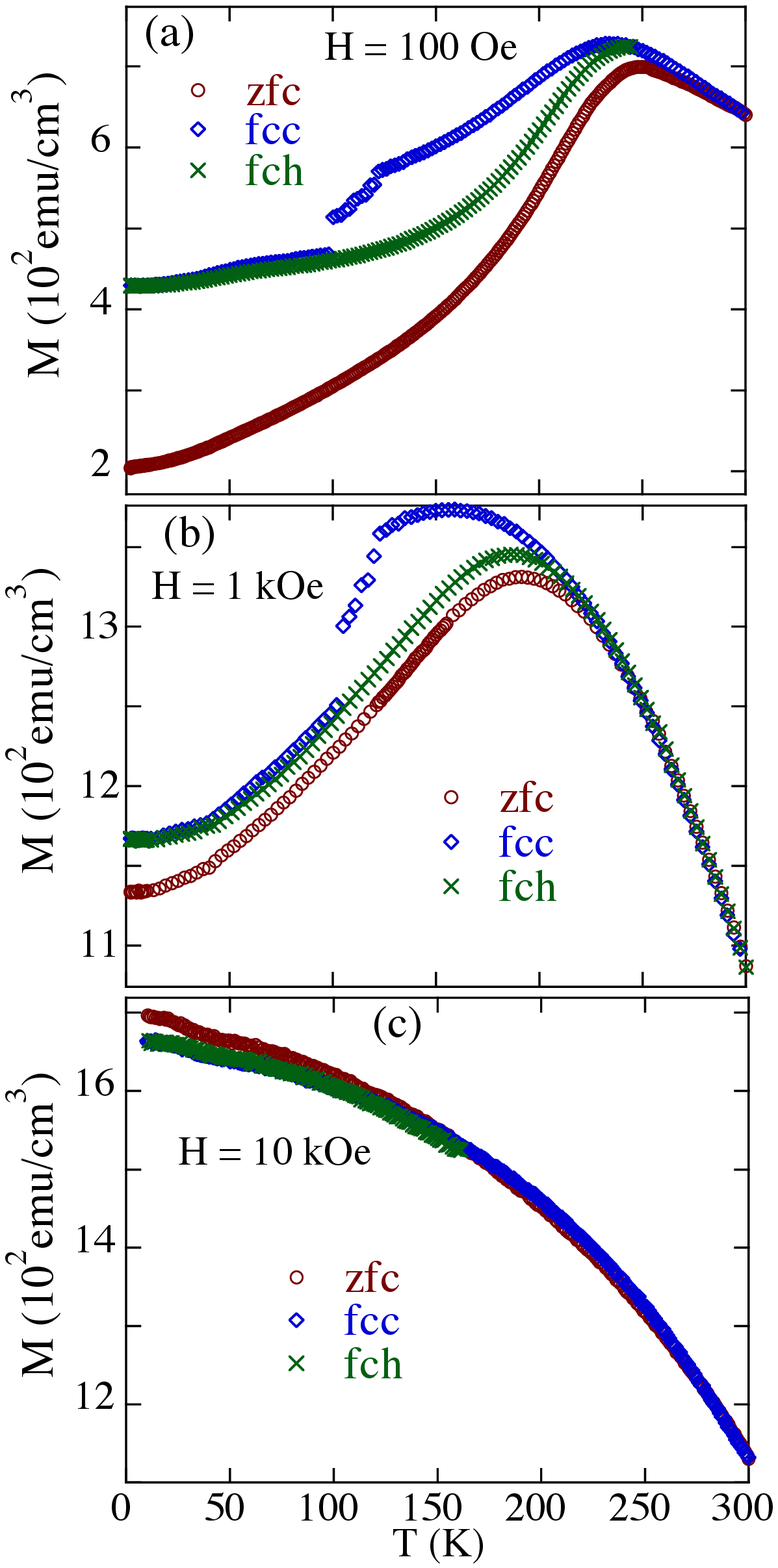}
\caption{Magnetisation curves as obtained for the FeRh-Fe$_3$Pt  film for: a) 100 Oe; b) 1 kOe; c) 10 kOe.}
\label{fig3}
\end{figure} 

Figure 3 shows isofield curves for  FeRh-Fe$_3$Pt sample with $H$ = 100 Oe (Fig. 3a), 1 kOe (Fig. 3b) and 10 kOe (Fig. 3c) (figure 3a is the same of Fig. 3 and it is included here for the sake of comparison). It is possible to see on Figures 3a and 3b, the anomaly in magnetisation occurring on the fcc curves, where $M$ starts dropping to lower values (with some apparent discontinuity) as temperature is cooled down below $\approx$ 120 K. The size, temperature, and width of the  anomaly in the fcc magnetisation curves do not show considerable changes with field. For $H$ = 10 kOe (Fig. 3c),  the anomalous effect is absent (it is suggestive that this high field overcomes the antiferromagnetic-ferromagnetic exchange coupling) and, interestingly, the zfc curve appears above the fcc and fch curves while the zfc curves in Fig. 3a and 3b  appears below the fcc and fch curves. These differences are probably related to the alignment of ferromagnetic domains, which depends on the strength of the magnetic field.  As already mentioned, it is interesting to note that, for temperatures above 120 K, the fcc curves in Figs. 3a and 3b lie well above the zfc and fch curves, though the fcc curves match almost perfectly the fch curves below 100 K. The later suggest the existence of some antiferromagnetic ordering (within the $FeRh$ thin layer) occurring below 100 K. One may understand this fact as a coexistence of two phases below 100 K, an antiferromagnetic phase due to the FeRh thin layer and a ferromagnetic phase due to the bulk Fe$_3$Pt, which is dominant.   

To check for possible effects in the temperature region around 100 K, we plot in Fig. 4a isothermal $M vs. H$ curves obtained from  80 K to 130 K. The main figure shows a 150 Oe window plot of  the original $M(H)$ curves presented in the inset. It is possible to see that each curve of Fig. 4a exhibit a typical ferromagnetic hysteresis behavior with a perfect symmetric coercive fields. It is also possible to see that the coercive field increases monotonically as temperature decreases.

One should note that  the isofields $M vs. T$ curves of Fig. 3 show a maximum in the magnetisation located approximately at 240-250 K. To check for possible effects related to this maximum on $M vs. H$ curves  we obtained few isothermals for temperatures around 250 K which are presented in Fig. 4b. The curves in Fig. 4b show a typical ferromagnetic hysteresis behavior with no apparent change in the magnetisation isothermals as the maximum in the $M vs. T$ curves, when $T$ $\approx$ 250 K, is crossed. 
We mention (not shown) that we check for a possible frequency dependence of the anomaly (metamagnetic-like transition) by measuring ac susceptibility curves for three different frequencies: 50, 500 and 2000 Hz; with an ac magnetic field of 1 Oe and a dc magnetic field of 20 Oe. The resulting curves of the susceptibility amplitude versus temperature did not show any frequency dependent effect. We also mention that we search for time relaxation effects when in the fcc curve near the anomalous transition above 120 K, but no magnetic relaxation was observed.
\begin{figure}[t]
\includegraphics[width=\linewidth]{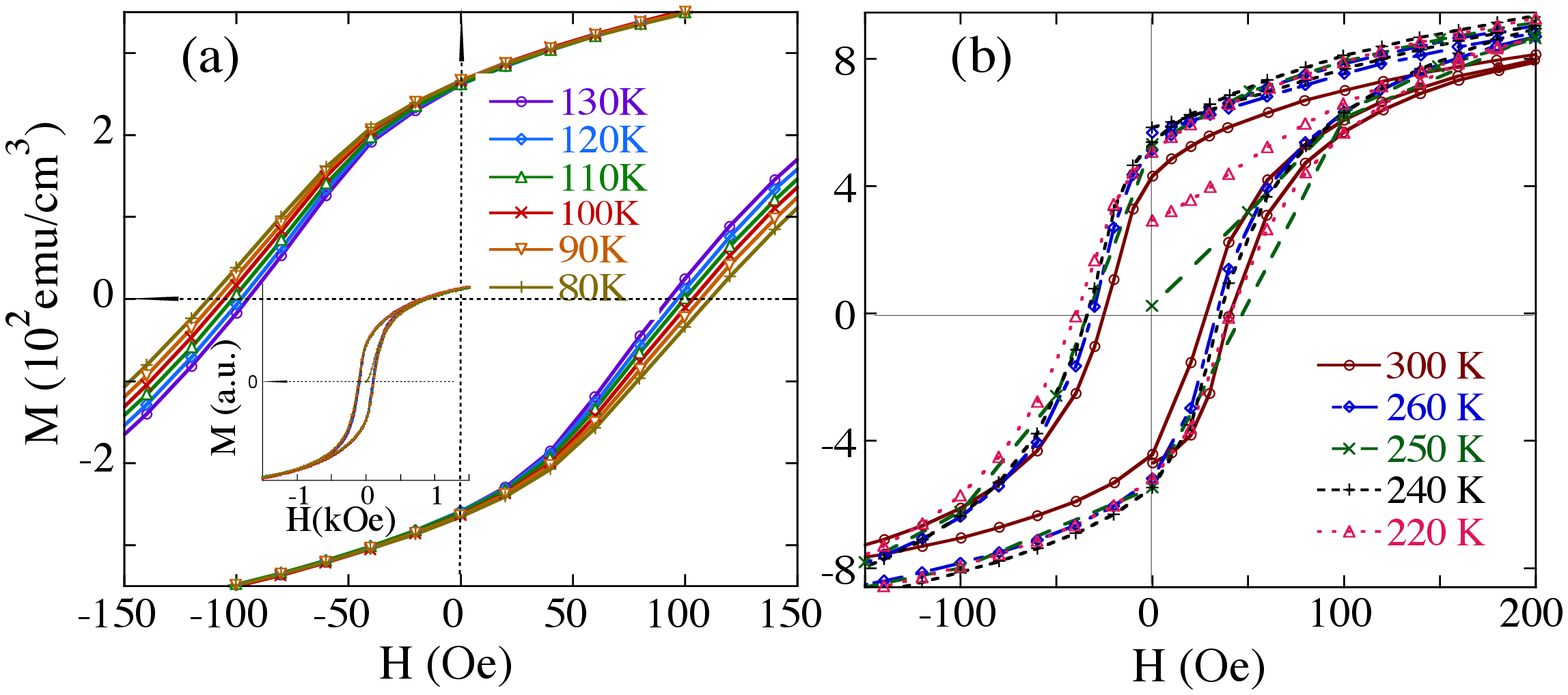}
\caption{a) Isothermal $M vs. H$ curves obtained at temperatures around 100 K plotted in a 150 Oe window. The inset shows the original $M vs. H$ curves; b) Isothermal $M vs. H$ curves obtained for several temperatures around 250 K.}
\label{fig4}
\end{figure} 

Although the metamagnetic-like transition observed here may have a different nature for bulk FeRh and thin films, it is interesting to compare the size of the jump occurring at $\approx$ 100 K.  From Fig. 3, the jump is estimated to be $\approx$ 5x10$^{3}$ emu/cm$^3$ for a 50 nm thickness while the size of the antiferromagnetic-ferromagnetic (AF-F) transition observed in FeRh in Ref.\onlinecite{maat} is of the order of 1.2x10$^{3}$ emu/cm$^3$, but at a $T$ $\approx$ 350 K. It should be mentioned that an AF-F transition was observed in a Fe$_{48}$Rh$_{52}$ film\cite{thiele} at $T$ $\approx$ 100 K, also with a size of the order of 1.2x10$^{3}$ emu/cm$^3$. So the "size" of the metamagnetic-like transition observed here for the FeRh-Fe$_3$Pt  film appears to be 4 times bigger than the AF-F transition observed for pure FeRh. This discrepancy might be also in part due to an under-estimative of our FeRh sample thickness. It should be mentioned that an increasing in the FeRh magnetisation above the antiferromagnetic transition has been observed in a FeRh/FePt thin film.\cite{thiele} 
 
 \begin{figure}[t]
\includegraphics[width=\linewidth]{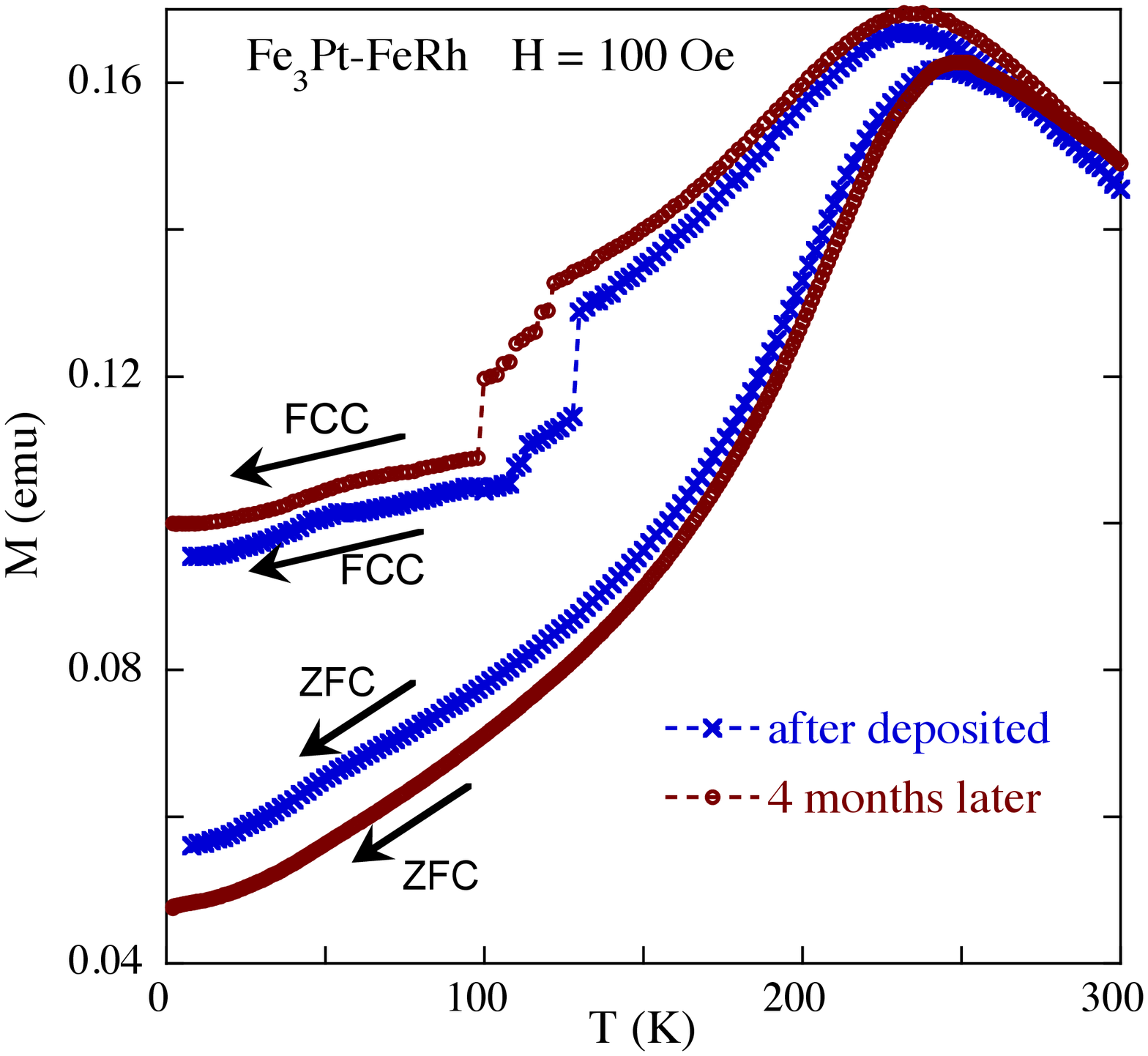}
\caption{Magnetisation curves as obtained for the  FeRh-Fe$_3$Pt film for 100 Oe with 4 months delay.}
\label{fig5}
\end{figure} 
Figure 5 depicts two measurements obtained in the same sample at same conditions but with approximately 4 months delay between each other. A comparison of both curves clearly shows the aging effect which shifted the onset temperature of the anomaly to a lower temperature ($\approx$ 20 degrees below), broadened its width and reduced the size of the apparent discontinuity (by $\approx$ 30$\%$). We lack, at moment, on an explanation for this aging effect. 

In conclusion we obtained a Fe$_{1-x}$Rh$_x$-Fe$_3$Pt with $x$ $\approx 0.5$ system by electrodeposition of FeRh on a Fe$_3$Pt ordered foil. The resulting film stack show, prior to further heat-treatments, an anomaly in the magnetisation when the sample is cooled in a magnetic field from 300 K down to lower temperatures, after reaching 300 K with the same applied field from a zero-field-cooled heating cycle from 1.8 K.  The anomalous magnetisation is reproducible and appears at temperatures close to 120 K, from fields as low as 20 Oe up to fields as high as 1 kOe, and resembles a metamagnetic transition. 

SSS, ADA, RBG, and FGG acknowledges support from CNPq and FAPERJ, Brazilian Agencies.

\end{document}